\documentclass[aps,prd,amsmath
,eqsecnum            
,preprintnumbers
,nofootinbib         
,twocolumn         
]{revtex4}
\usepackage[dvips]{graphicx}
\usepackage{amsmath}
\usepackage{amsfonts}
\usepackage{amssymb}
\usepackage{longtable}   

\allowdisplaybreaks[4]     

\begin{document}

\title{Szekeres Swiss-Cheese model and supernova observations}

\author{Krzysztof Bolejko}
\affiliation{N. Copernicus Astronomical Center, Polish Academy of Sciences, \\
Bartycka 18, 00 716 Warszawa, Poland}
\email{bolejko@camk.edu.pl}

\author{Marie-No\"elle C\'el\'erier}
\affiliation{Laboratoire Univers et Th\'eories (LUTH), \\
Observatoire de Paris, CNRS, Universit\'e Paris-Diderot \\
5 place Jules Janssen, 92190 Meudon, France}
\email{marie-noelle.celerier@obspm.fr}

\date { }

\begin{abstract}
We use different particular classes of axially symmetric Szekeres Swiss-cheese models for the study of the apparent dimming of the supernovae of type Ia. We compare the results with those obtained in the corresponding Lema\^itre--Tolman Swiss-cheese models. Although the quantitative picture is different the qualitative results are comparable, i.e, one cannot fully explain the dimming of the supernovae using small scale ($\sim 50$ Mpc)
inhomogeneities. To fit successfully the data we need structures 
of order of 500 Mpc size or larger. However, this result might be an artifact due to the use of axial light rays in axially symmetric models. Anyhow, this work is a first step in trying to use Szekeres Swiss-cheese models in cosmology and it will be followed by the study of more physical models with still less symmetry.
\end{abstract}

\maketitle


\section{Introduction}
The Universe, as we observe it, is inhomogeneous almost at all scales. However, the standard approach, based on the cosmological principle, assumes that homogeneous models with Robertson-Walker (RW) geometry can be successfully employed in a cosmological framework, i.e., above some $\sim$100 Mpc scale. However, within such a framework, one needs to assume that the Universe is filled with dark energy in order to achieve concordance with SNe Ia observations. But dark energy has never been directly observed neither in the Universe, nor in laboratories, and it has very exotic properties, namely, it is a kind of gas with negative pressure or a mere cosmological constant with an amplitude too small to account for the vacuum energy in the standard model of particle physics. Therefore, it is of the utmost importance to test whether the observations can be explained with models that do not require dark energy but that instead take inhomogeneities into account.

A number of approaches have been suggested so far. Among them the one patch spherically symmetric Lema\^itre--Tolman models with a central observer \cite{celerier00, iguchi02, alnes06, chung06, alnes07, enqvist07, ABNV09, bolejko08a, garciabellido08, garciabellido08b,zibin08,yoo08,enqvist08, BW09,CBKH09},
the thin shell approximations \cite{tomita00,tomita01a,tomita01b,CS08}, the Stephani models \cite{dabrowski98,godlowski04,stelmach06,dabrowski07},
and the Szekeres models \cite{MI1,MI2}. However, all these approaches have their restrictions.

For example, for the Lema\^itre-Tolman model assuming spatial spherical symmetry around the observer, this symmetry has been turned into an argument stating that the observer's location at or near the symmetry center is too special as regards the requirement of the cosmological principle. Now, we want to stress here that these models should not be considered as assuming we are living at or near the center of any spherically symmetric universe. They should merely be considered as a first step in the process of modeling cosmological inhomogeneities, i.e., as a mere smoothing out of the inhomogeneities on angular scales, analogous to their smoothing out on the totality of space in homogeneous models. Thus, the subsisting inhomogeneities are only radial. The use of such models must be merely regarded as a first approach which will be followed by more precise ways of dealing with the observed inhomogeneities.
The Stephani class of solutions are conformally flat \cite{Krasinski97} and cannot therefore constitute general models for the Universe. The thin shell approximations are based on two RW models and are also spherically symmetric around the observer.
The Szekeres models are the most general ones, 
even if the five functions defining a model 
depend only on the radial coordinate.

An improvement is to use Swiss-Cheese models. Some models of this class where the holes were Lema\^itre--Tolman patches have already been proposed in the literature \cite{BTT07,MKMR07,ABNV09,BN08,BTT08,KMM09}. However, the results suggest that
we need at least structures of several hundred Mpc in order to reproduce the supernova data. In Ref. \cite{CF09} the model proposed contains Schwarzschild regions,
each with the mass of a galaxy. The results suggest that such a geometry cannot explain dark energy as an effect of inhomogeneities.

We propose here a generalization of these approaches, where instead of Lema\^itre--Tolman or Schwarzschild patches, which are all spherically symmetric, we use Szekeres models to represent the holes in the homogeneous cheese. The Szekeres model has several advantages, like anisotropic density distribution which leads to a different evolution pattern than in the Lema\^itre--Tolman model \cite{B06,B07}. Thus there is a possibility that the Szekeres Swiss-cheese models can come up with a more realistic picture, where the inhomogeneities are no more spherical and therefore can give a more physical representation of our Universe. However, for calculation simplicity and as a first step, we limit ourselves here to the study of an axially symmetric propagation of light through axially symmetric Szekeres holes.

The structure of the present paper is as follows. 
Section \ref{sec3} is devoted to the presentation of the Szekeres models and of the particular subclasses used in this paper. In Sec.~\ref{sec4}, we give the residual Hubble diagrams for the five axially symmetric quasi-spherical Szekeres Swiss-cheeses considered here and we compare them to the corresponding Lema\^itre--Tolman models, to the $\Lambda$CDM model and to the actual data. In Sec.~\ref{sec5}, we display our conclusions.

\section{Szekeres models} \label{sec3}

The Szekeres solutions \cite{S75} are the most general solutions of Einstein's equations one can obtain with a dust gravitational source. They have no symmetry, i.e., no Killing vector, and are therefore well-suited to describe a lumpy universe. Their metric in comoving coordinates and synchronous time gauge is

\begin{equation}
{\rm d} s^2 =  c^2 {\rm d} t^2 - {\rm e}^{2 \alpha} {\rm d} r^2 - {\rm e}^{2 \beta} ({\rm d}x^2 + {\rm d}y^2),
\label{metsz}
\end{equation}
where $\alpha$ and $\beta$ are functions of $(t,x,y,r)$ to be determined by the field equations.

There are two families of Szekeres solutions. The first family, where $\beta' = 0$ (here the prime denotes derivation as respect to $r$) is a simultaneous generalization of the Friedmann and Kantowski-Sachs models. Since it has found so far no application in cosmology, we do not discuss it here. The second family of solutions is obtained when $\beta' \neq 0$. When the Einstein equations are solved, its metric can be written, after a change of coordinates more convenient for our purpose \cite{H96},

\begin{equation}
{\rm d} s^2 =  c^2 {\rm d} t^2 - \frac{(\Phi' - \Phi {  E}'/ {  E})^2}
{\epsilon - k} {\rm d} r^2 - \Phi^2 \frac{({\rm d} x^2 + {\rm d} y^2)}{{  E}^2},
\label{metsz2}
 \end{equation}
where $\epsilon = 0, \pm 1$, $\Phi$ is a function of $t$ and $r$ and $k$ is a function of $r$.
\begin{equation}
{  E} = \frac{S}{2} \left[ \left( \frac{x-P}{S} \right)^2
+ \left( \frac{y-Q}{S} \right)^2 + \epsilon \right],
\label{Edef}
\end{equation}
with $S(r)$, $P(r)$, $Q(r)$, functions of $r$.

\subsection{Quasi-spherical Szekeres models}

As can be seen from eq. (\ref{metsz2}), only $\epsilon = +1$ allows the solution to have all the three Friedmann limits (hyperbolic, flat and spherical). This is induced by the requirement of the Lorentzian signature of the metric. Since we are interested in the Friedmann limit of our model which we expect to become homogeneous at very large scales, i.e., that of the last-scattering, we will focus only on the $\epsilon = +1$ case. Such a case is called the quasi-spherical Szekeres model.

Its metric, obtained for $\epsilon = +1$, becomes

\begin{equation}
{\rm d} s^2 =  c^2 {\rm d} t^2 - \frac{(\Phi' - \Phi {  E}'/ {  E})^2}
{1 - k} {\rm d} r^2 - \Phi^2 \frac{({\rm d} x^2 + {\rm d} y^2)}{{  E}^2},
\label{ds2}
 \end{equation}
where
\begin{equation}
{  E} = \frac{S}{2} \left[ \left( \frac{x-P}{S} \right)^2
+ \left( \frac{y-Q}{S} \right)^2 +1 \right].
\end{equation}
In the spherical coordinates, ${  E} = S/(1-\cos\vartheta)$.

Applying the Einstein equations to the metric (\ref{ds2}) and assuming the energy momentum tensor is that of dust, the Einstein equations reduce to the following two:

\begin{equation}
\frac{1}{c^2}\dot{\Phi}^2 = \frac{2M}{\Phi} - k + \frac{1}{3} \Lambda
\Phi^2, \label{vel}
\end{equation}
where $M(r)$, is an arbitrary function related to the density $\rho$ via:
\begin{equation}
\kappa \rho c^2= 
 \frac{2M' - 6 M {  E}'/{  E}}{\Phi^2 ( \Phi' - \Phi {  E}'/{  E})}. \label{rho}
\end{equation}
where $\kappa=8\pi G/c^4$, $\Lambda$ is the cosmological constant and 
$M$ is an arbitrary function of $r$. 
The 3D Ricci scalar is
\begin{equation}
^{3}\mathcal{R} = 2 \frac{k}{\Phi^2} \left( \frac{ \Phi k'/k - 2 \Phi
{  E}'/{  E}}{ \Phi' - \Phi {  E}'/{  E}} + 1 \right).
\label{3dr}
\end{equation}
The Weyl curvature decomposed into its electric and magnetic part is
\begin{eqnarray}\label{Weyl}
&& E^{\alpha}{}_{\beta} = C^{\alpha}{}_{\gamma \beta \delta} u^{\gamma} u^{\delta} =
\frac{M(3 \Phi' - \Phi M'/M)}{3 \Phi^3 ( \Phi' - \Phi E' / E)}
{\rm diag} (0,2,-1,-1), \nonumber \\
&& H_{\alpha \beta} = \frac{1}{2} \sqrt{-g} \rho_{\alpha \gamma \mu \nu} C^{\mu
\nu}{}_{\beta \delta} u^{\gamma} u^{\delta} = 0.
\end{eqnarray}

As in the Lema\^itre--Tolman model, the bang time function, $t_B(r)$, follows from
(\ref{vel}):
\begin{equation}
\int\limits_0^{\Phi}\frac{{\rm d} \widetilde{\Phi}}{\sqrt{- k + 2M /
\widetilde{\Phi} + \frac 1 3 \Lambda \widetilde{\Phi}^2}} = c [ t - t_B(r)].
\label{tbf}
\end{equation}

Since all the formulae given so far are covariant under coordinate transformations of the form $\tilde{r} = g(r)$, this means that one of the functions $k(r)$, $S(r)$, $P(r)$, $Q(r)$, $M(r)$ and $t_B(r)$ can be fixed at our convenience by the choice of $g$. Hence, each Szekeres solution is fully determined by only five functions of $r$: in the following, we choose $k$, $S$, $P$, $Q$ and $M$.

\subsection{Axially symmetric quasi-spherical Szekeres models}

As a first step, and for simplification purpose, we consider here axially symmetric quasi-spherical Szekeres models. Actually, since in our Swiss-cheese model described below, we use only radially directed light rays, it has been shown in \cite{ND07} (see also \cite{BKHC09}) that this implies that the Szekeres model should be axially symmetric. The simplest axially-symmetric Szekeres model obeys 

\[ P(r) = x_0 = {\rm ~const}, \quad \quad Q(r) = y_0 = {\rm ~const}. \]
In this case the dipole axis is along $x=x_0$ and $y=y_0$ (or in spherical coordinates along the directions $\vartheta=0$ and $\vartheta=-\pi$).

For the axially directed geodesics (${\rm d} x = {\rm d} y = 0$), we obtain from (\ref{ds2})

\begin{equation}
 \frac{{\rm d} t}{{\rm d} r} = \pm \frac{1}{c} \frac{\Phi' - \Phi {  E}'/{  E}}{\sqrt{1 - k}}.
\label{snge}
\end{equation}
The plus sign is for $r_e < r_o$ and the minus sign for $r_e > r_o$, with $r_e$, the radial coordinate of the source and $r_o$, the radial coordinate of the observer.

The redshift relation in this case is \cite{BKHC09}:

\begin{equation}
\ln (1+z) = \pm \frac{1}{c} \int\limits_{r_e}^{r_o} {\rm d} r
\frac{ \dot{\Phi}' - \dot{\Phi} {  E}'/{  E}}{\sqrt{1 - k}},
\label{srf}
\end{equation}
or equivalently:
	
\begin{eqnarray}
 \frac{{\rm d} r}{{\rm d} z} = \pm \frac{c}{1+z} \frac{\sqrt{1-k}}{
\dot{\Phi}' - \dot{\Phi} {  E}'/{  E}}, \nonumber \\
 \frac{{\rm d} t}{{\rm d} z} = \frac{1}{1+z} \frac{\Phi' - \Phi {  E}'/{  E}}{
\dot{\Phi}' - \dot{\Phi} {  E}'/{  E}}.
\label{redrel}
\end{eqnarray}

\subsection{The Swiss Cheese models}

As seen above, in the case of axially directed geodesics the equations which describe light propagation simplify significantly.
Moreover, density fluctuations (\ref{rho}) and curvature fluctuations [eqs. (\ref{3dr})
and (\ref{Weyl})]
are the largest along the axial axis ($\pm {  E}'/{  E}$ - 
when a light ray passes through the origin
${  E}'/{  E} \rightarrow -{  E}'/{  E}$, since
$E'/E$ is always multiplied by $\Phi$ or $\dot{\Phi}$ which are zero at the
origin thus there is no discontinuity here.).
Therefore in constructing our Swiss-cheeses models we focus only on axial geodesics.

When constructing a Swiss Cheese model, we need to satisfy the junction
conditions for matching. Here we match the Szekeres inhomogeneity (holes)
to the Friedmann background (cheese).
These Szekeres patches are placed so that
their boundaries touch wherever a light ray exits one inhomogeneous patch.
Thus the ray immediately enters another Szekeres inhomogeneity and does not 
spend any time in the Friedmann background. 
To match a Szekeres patch to a Friedmann background across a comoving
spherical surface, $r =$ constant, the conditions are: that the mass inside the
junction surface in the Szekeres patch is equal to the mass that would be inside
that surface in the homogeneous background;
that the spatial curvature  at the junction surface is the same in both
the Szekeres and Friedmann models, which implies that $k_{SZ} = k_F r^2$
and $(k_{SZ})' = 2 k_F r$; finally that the bang time and also $\Lambda$ must be continuous across the junction -- 
unless otherwise stated we  assume that $\Lambda=0$.

Besides matching the inhomogeneous patches we also need to take care
of the null geodesics. However,
as in this case we only consider
axial geodesics the junction is trivial and requires only
matching the radial, or equivalently the time component \cite{KB09}.

We consider 3 different types of models with 3 different backgrounds.
The first model consists of small-scale inhomogeneities 
within an open FLRW model ($\Omega_m = 0.25$).
The others have larger patches within either an Einstein-de Sitter model or
an open FLRW model with $\Omega_m = 0.9$.

Light propagation is obtained by solving eqs. (\ref{redrel}).
The luminosity distance is calculated as follows.
First we solve for the angular diameter distance, $D_A$,
which is defined as \cite{E71}:

\begin{equation}
D_A^2 : = \frac{\delta S}{\delta \Omega},
\label{dadef}
\end{equation}
where $\delta S$ is the cross-sectional area of the bundle of null geodesics diverging from a radiation source, perpendicular to the propagation vector of light at a point with affine parameter $s$ and $\delta \Omega$ is the solid angle subtending this area.
The rate of change of $\delta S$ is given by \cite{S61}:

\begin{equation}
\frac{ {\rm d} \delta S}{ {\rm d} s} = 2 \theta \delta S,
\label{schd}
\end{equation}
where $\theta$ is the expansion of the family of null geodesics.
Using the Sachs propagation equations \cite{S61,SEF92}

\begin{eqnarray}
\frac{{\rm d} \theta}{{\rm d} s} + \theta^2 + |\sigma|^2 = 
- \frac{1}{2} R_{\alpha \beta} k^{\alpha} k^{\beta}, \label{otevo} \\
\frac{{\rm d} \sigma}{{\rm d} s} + 2 \theta \sigma = 
- \frac{1}{2} R_{\alpha \beta \mu \nu} \epsilon^{*\alpha}
 k^{\beta} \epsilon^{*\mu} k^{\nu}, \label{sgmevo}
\end{eqnarray}
where $R_{\alpha \beta \mu \nu}$ is the Riemann tensor, $R_{\alpha \beta}$ is the Ricci tensor, $|\sigma|^2 = (1/2) \sigma_{\alpha \beta} \sigma^{\alpha \beta}$, and a star
denotes a complex conjugate. 
For axial geodesics $\epsilon^\alpha = E (\delta^\alpha{}_2 + i \delta^\alpha{}_3) / (\sqrt{2} \Phi)$.
Which implies that if initially $\sigma =0$ then it will remain zero.
Therefore, the relation for the angular diameter
distance is 

\begin{equation}
\frac{{\rm d^2} D_A}{{\rm d} s^2}  = - \frac{1}{2} R_{\alpha \beta} k^{\alpha} k^{\beta} D_A.
\label{darel}
\end{equation}

Finally using the reciprocity theorem \cite{Eth33,E71} the luminosity
distance is
\begin{equation}
D_L = D_A (1+z)^2.
\end{equation}

\section{Results}\label{results} \label{sec4}

\begin{figure*}
\begin{center}
\includegraphics[scale=0.56]{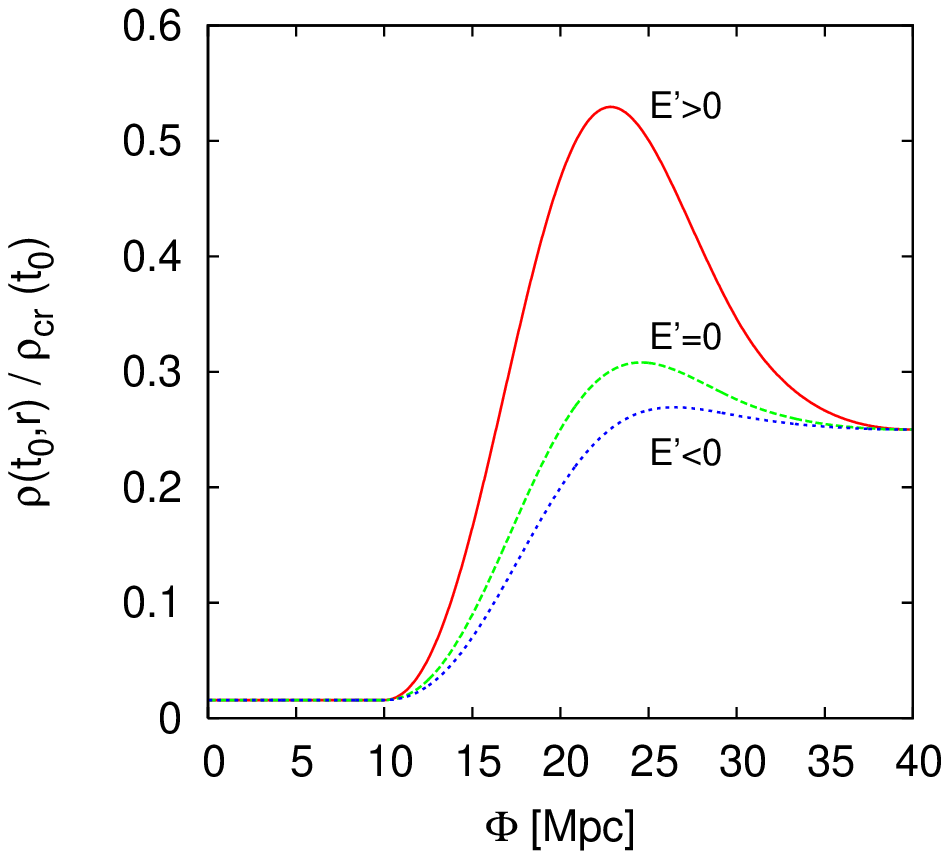}
\includegraphics[scale=0.56]{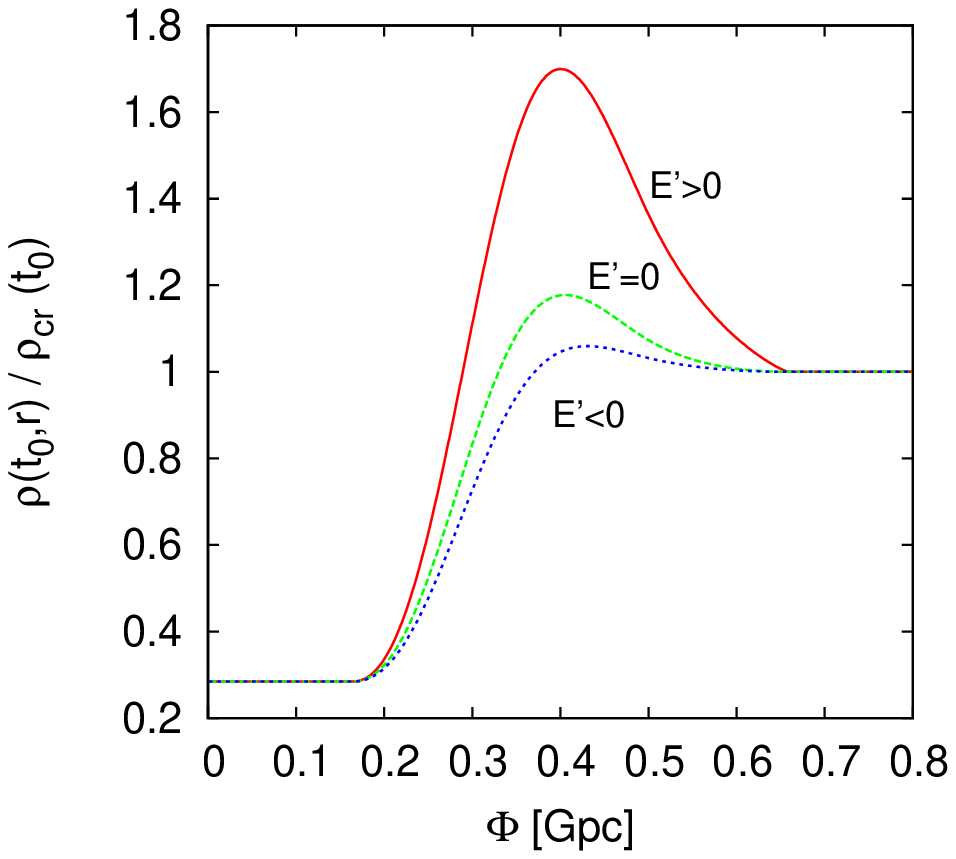}
\includegraphics[scale=0.56]{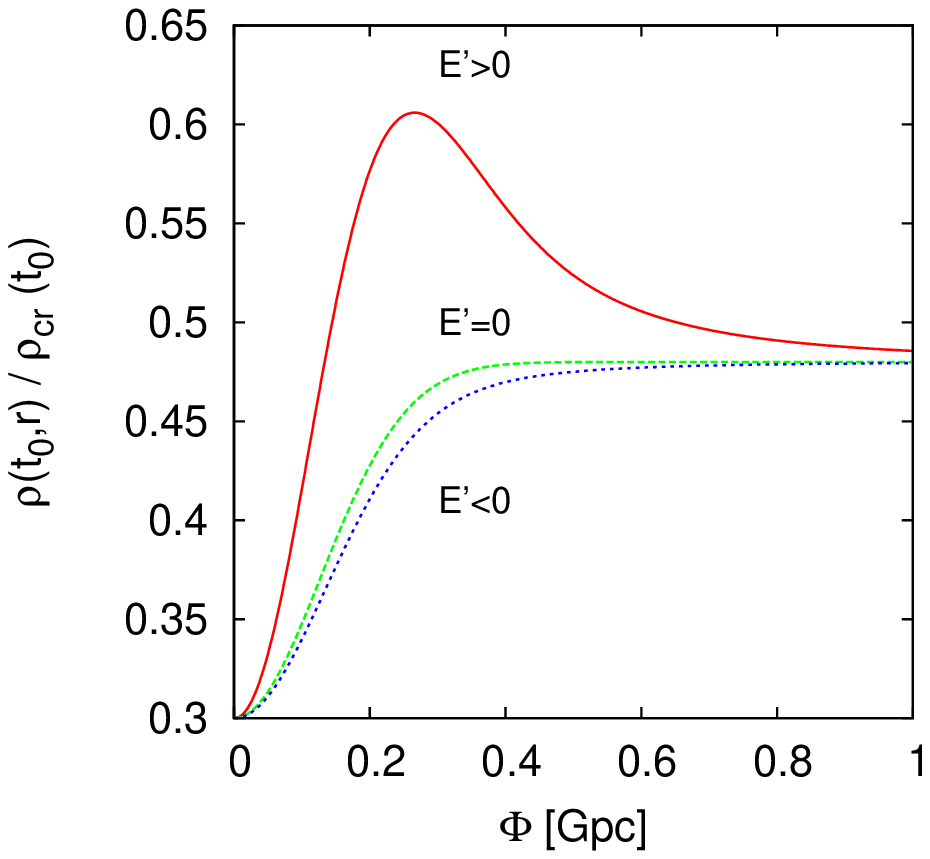}
\caption{Density profiles along the axial geodesic for the models studied in this paper ---
model 1 (left), models 2 and 3 (center), models 4 and 5 (right).}
 \label{fig0}
\end{center}
\end{figure*}

\subsection{Small scale ($\sim$ 50 Mpc) inhomogeneities}\label{ssinh}

We first study a Swiss-cheese which can be considered as representing our local Universe with voids of size around 50 Mpc \cite{jon09}. Let us consider the Swiss-cheese model exhibiting holes each described by Szekeres patches whose 5 arbitrary functions, $M$, $k$, $Q$, $P$, and $S$ are of the following forms:

\begin{equation}\label{Mm1}
M = M_0 +  \left\{ \begin{array}{ll}
M_1 \ell^3 & {\rm ~for~} \ell \leqslant x_a, \\
M_2 \exp \left[ -  \left( \frac{ \ell - 2x_a}{x_a} \right)^2 \right] & {\rm ~for~} x_a \leqslant  \ell \leqslant 3x_a \\ 
-M_1 (\ell - 4x_a)^3 & {\rm ~for~} 3x_a \leqslant  \ell \leqslant  4x_a,  \\
0 &  {\rm ~for~} \ell \geqslant 4x_a, 
\end{array} \right.
\end{equation}
where $\ell = r$/kpc, $M_0 = (4 \pi G /3c^2) \rho_{b} \ell^3$,
$\rho_{b} =  \Omega_m \frac{3H_0^2}{8 \pi G}$,
$\Omega_m = 0.25$, $H_0=72$ km s$^{-1}$ Mpc$^{-1}$,
$x_a= 10^4$,
$M_1 = x_a^{-3} M_2 {\rm e}^{-1.5}$, $M_2= -7 \times 10^{11}$ kpc. 
The above profile was chosen for the following reasons:
it behaves like a FLRW model for $\ell \leqslant x_a$ but with lower
density than outside, then for $x_a \leqslant  \ell \leqslant 3x_a$
we have a transition region, and a cubic behavior for $ 3x_a \leqslant  \ell \leqslant  4x_a$,
which allows for a smooth matching to the background values. 
Although the distribution of mass, i.e. the function $M(r)$
is spherically symetric (it only depends on $r$), the density distribution,
as seen from \ref{rho}, is not.
For the same reason we chose the following profile for the function $k$: 
\begin{equation}\label{km1}
k = k_0- \frac{1}{2} \times  \left\{ \begin{array}{ll}
k_1 \ell^2 & {\rm ~for~} r \leqslant x_b, \\
k_2 \exp \left[ - \left( \frac{ \ell - x_b}{x_b} \right)^2 \right] & {\rm ~for~} x_b
 \leqslant  r \leqslant 3 x_b \\
k_1 (4 x_b - \ell)^2 & {\rm ~for~} 3 x_b \leqslant  r \leqslant  4 x_b,  \\
0 &  {\rm ~for~} r \geqslant 4x_b,
\end{array} \right.
\end{equation}
where $k_0 = (\kappa \rho_b c^2/3 - H_0^2)\ell^2$,
$k_1 = k_2 x_b^{-2} {\rm e}^{-1}$, $k_2 = -8.84 \times 10^{-6}$,
$x_b= 8.68 \times 10^3$.
\begin{eqnarray}
&& S = (5 \times 10^3 + \ell)^{\pm 0.78} \\
&& P = 1= x_0, \\
&& Q = 1 =y_0.
\end{eqnarray}
where $+$ is for propagation from the origin
[${  E}'/{  E} = 0.78/(5 \times 10^3 + \ell)$], and $-$ towards the origin
[${  E}'/{  E} = -0.78/(5 \times 10^3 + \ell)$].
This model will be referred to as model 1.
As can be seen from (\ref{rho})--(\ref{Weyl}) and (\ref{Mm1})--(\ref{km1}),
for $r>40$ Mpc model 1 becomes the homogeneous FLRW model.
The density profile along the axial geodesic is presented in the left panel of Fig. \ref{fig0}. First light propagates towards the center, $E'>0$, and after passing through the origin, $E'$ becomes negative, and so on.

In model 1 we place the observer in the homogeneous region where
$r_o=40$ Mpc. We join Szekeres patches at $r=40$ Mpc.
Thus Szekeres patches are placed so that
their boundaries touch and wherever a light ray exits one inhomogeneous patch
it immediately enters another Szekeres inhomogeneity and does not spend any time in
the FLRW background.

\begin{figure*}
\includegraphics[scale=0.8]{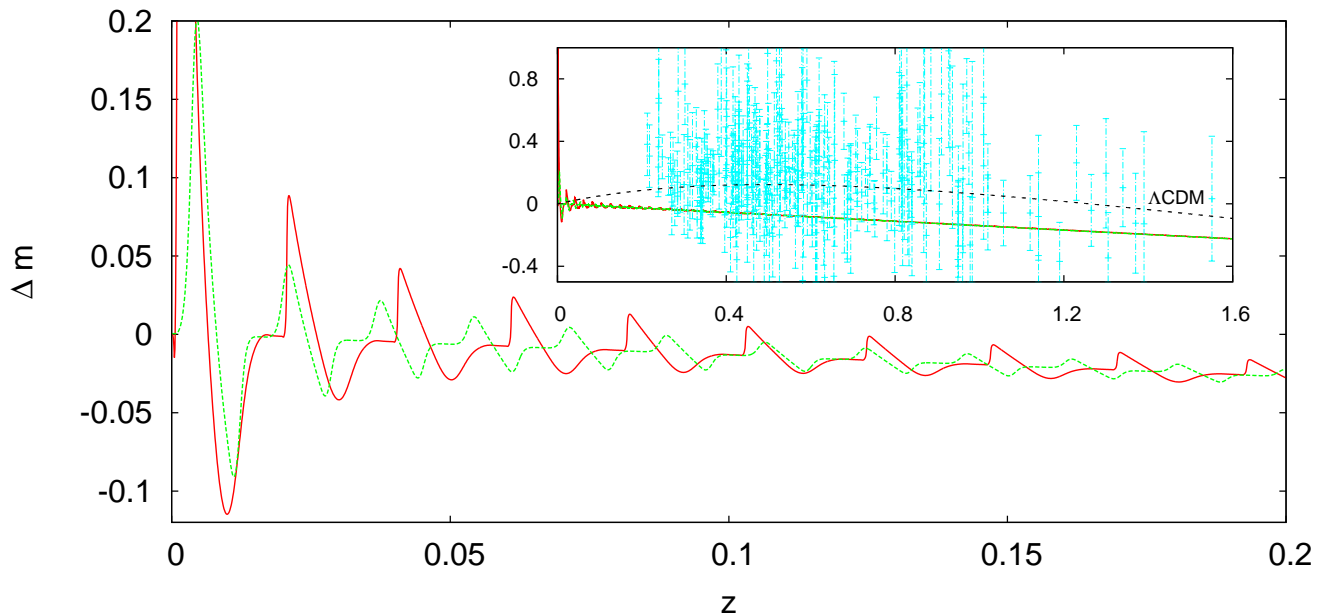}
\caption{The residual Hubble diagram for model 1 (red). For comparison, the
results for the corresponding Lema\^itre--Tolman model (i.e. with $M$ and $k$ as in model 1 but with $E' = 0$) are also presented (green). The Union supernova data set \cite{K08} is depicted in the inset, for clarity only measurements for $z>0.2$ are given.}
\label{fig1}
\end{figure*}

The results for model 1 in the form of a residual Hubble diagram  are presented in Fig. \ref{fig1}.
The residual Hubble diagram presents $\Delta m$ as a function of redshift:
\[\Delta m = m - m^{emp} = 5 \log \frac{D_L}{D^{emp}_L}.\]
where $m^{emp}$ and $D^{emp}_L$ are the expected magnitude and luminosity distance in an empty FLRW model.
 As seen
in comparison with the corresponding Lema\^itre--Tolman Swiss-cheese model the peaks are of larger amplitude. However, for $z>0.2$ the fluctuations  of the magnitude are negligible and the results are almost the same as in the hyperbolic FLRW background model.
Thus as previously noted \cite{BTT07, BTT08, KMM09} the Lema\^itre--Tolman Swiss-cheese models with small scale inhomogeneities are not sufficient to explain away dark energy, and this is also the case for this particular axially symmetric quasi-spherical Szekeres Swiss-cheese.

\subsection{Medium scale ($\sim$ 500 Mpc) models with homogeneous center}\label{ssinh}

The Swiss-cheese model considered in this section possesses holes that are described by the following functions:

\begin{equation}
M = M_0 +  \left\{ \begin{array}{ll}
M_1 \ell^3 & {\rm ~for~} \ell \leqslant x_a, \\
M_2 \exp \left[ - 3 \left( \frac{ \ell - 2x_a}{x_a} \right)^2 \right] & {\rm ~for~} x_a \leqslant  \ell \leqslant 3x_a \\ 
-M_1 (\ell - 4x_a)^3 & {\rm ~for~} 3x_a \leqslant  \ell \leqslant  4x_a,  \\
0 &  {\rm ~for~} \ell \geqslant 4x_a, 
\end{array} \right.
\end{equation}
where $\ell = r$/kpc, $M_0 = (4 \pi G /3c^2) \rho_{b} \ell^3$,
$\rho_{b} =  \frac{3H_0^2}{8 \pi G}$, $H_0=72$ km s$^{-1}$ Mpc$^{-1}$,
$x_c= 1.65 \times 10^5$,
$M_1 = x_a^{-3} M_2 {\rm e}^{-1.5}$, $M_2= -2.4 \times 10^{15}$ kpc, 
\begin{equation}
k = k_0- \frac{1}{2} \times  \left\{ \begin{array}{ll}
k_1 \ell^2 & {\rm ~for~} r \leqslant x_d, \\
k_2 \exp \left[ - \left( \frac{ \ell - x_d}{x_d} \right)^2 \right] & {\rm ~for~} x_d
 \leqslant  r \leqslant 3 x_d \\
k_1 (4x_d - \ell)^2 & {\rm ~for~} 3 x_d \leqslant  r \leqslant  4 x_d,  \\
0 &  {\rm ~for~} r \geqslant 4x_b,
\end{array} \right.
\end{equation}
where $k_0 = (\kappa \rho_b c^2/3 - H_0^2)\ell^2$,
$k_1 = k_2 x_d^{-2} {\rm e}^{-1}$, $k_2 = -5.05 \times 10^{-3}$
$x_d = 1.46 \times 10^5$,
\begin{eqnarray}
&& S = {\rm e}^{\pm \varepsilon 1.5 \times 10^{-8} r} \\
&& P = 1 = x_0, \\
&& Q = 1= y_0.
\end{eqnarray}
where $+$ is for propagation from the origin
[${  E}'/{  E} =  \varepsilon 1.5 \times 10^{-8}$],
and $-$ towards the origin
[${  E}'/{  E} = - \varepsilon 1.5 \times 10^{-8}$].
We consider two models: model 2 with 
$\varepsilon=1$ and model 3 with $\varepsilon=-1$.
As can be seen from (\ref{rho})--(\ref{Weyl})
for $r>660$ Mpc these models become the Einstein-de Sitter model.
Also, from the construction it can be seen that for $r<x_d = 146$ Mpc these models are homogeneous.

In models 2 and 3 we place the observer at the origin,
where $\rho/\rho_{cr} \approx 0.28$.
We join Szekeres patches at $r=660$ Mpc.
As in the previous case the Szekeres patches are placed so that
their boundaries touch.
Since we place the observer at the origin, we have two choices:
we can send a light ray towards the largest dipole fluctuations ($E'>0$) -- model 2 -- or in an opposite direction ($E'<0$) -- model 3.

\begin{figure*}
\includegraphics[scale=0.8]{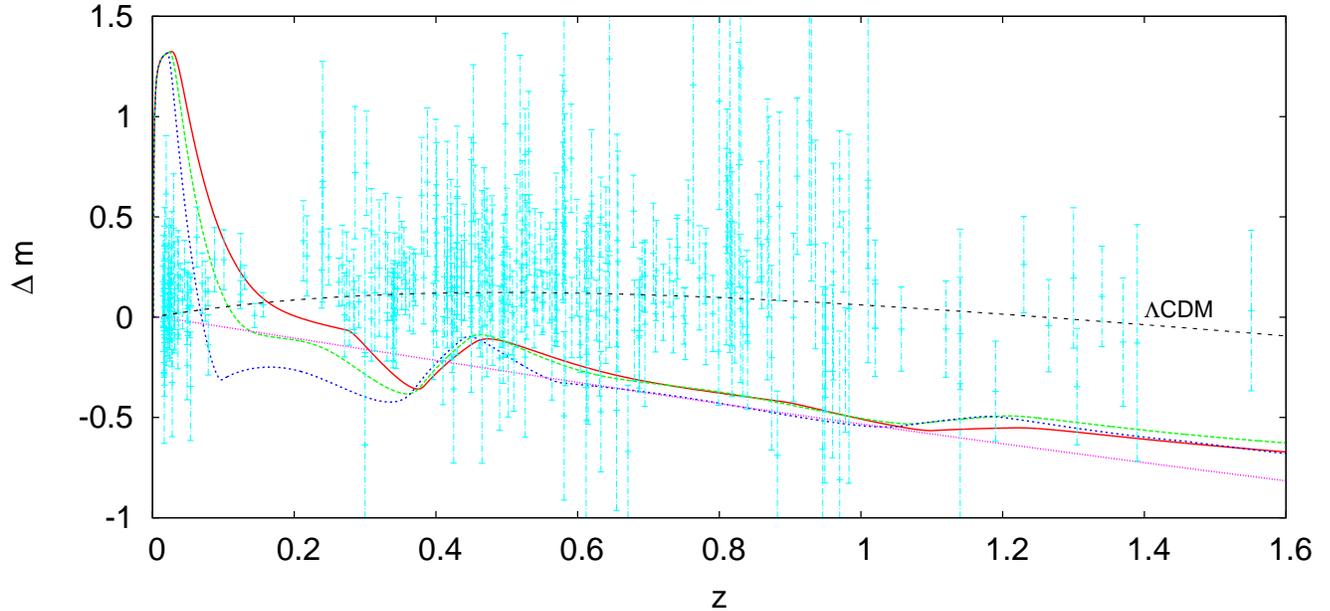}
\caption{The residual Hubble diagram for model 2 (blue), 
model 3 (red) and for the corresponding Lema\^itre--Tolman model (i.e. with $M$ and $k$ as in models 2-3 but with $E' = 0$). Supernovae measurements are extracted from the Union data set \cite{K08}.}
\label{fig2}
\end{figure*}

The results are presented in Fig. \ref{fig2}. 
As can be seen, for $z>0.6$ the magnitude fluctuations have small amplitudes. 
Moreover, the first peak is undesirable, since it omits all low-z supernovae,
and, since low-z measurements have the lowest errors,
the $\chi^2$ is very high and is equal to $2419.79$
for model 3 and $2291$ for model 2.
For 307 supernova measurements this is an extremely high value\footnote{$\chi^2 = \sum_i \frac{(\mu_i - \mu_0)^2}{\sigma_i^2 + \sigma_{int}^2}$
where $\mu_i$ and $\sigma_i$ correspond to the measurements of the 307 supernovae \cite{K08}, $\mu_0$ is the distance modulus in the empty FLRW model, and $\sigma_{int} =0.12$.} . As in the previous case the results are qualitatively comparable with those reported when using the Lema\^itre--Tolman Swiss-cheese models \cite{BN08,MKMR07}
and show that these types of models are not suitable for fitting supernova data.

\subsection{Medium scale ($\sim$ 500 Mpc) models with inhomogeneous center
and the minimal void scenario}\label{ssinh}

The Swiss-cheese model considered in this section exhibits holes that are described by the following functions:

\begin{equation}
M = \frac{\rho_b \kappa c^2}{2} \int\limits_0^r {\rm d} \tilde{r} \, \tilde{r}^2  \left[ 1 + \delta_{\rho} - \delta_{\rho} \exp \left( -
\frac{\tilde{r}^2}{\sigma^2} \right) \right],
\label{mas5}
\end{equation}
where 
$\rho_{b} =  \Omega_m \frac{3H_0^2}{8 \pi G}$,
$\Omega_m = 0.3$, $H_0=68$ km s$^{-1}$ Mpc$^{-1}$,
$\delta_{\rho} = 0.6$ and $\sigma = 180$ Mpc.
The function $k$ is calculated from (\ref{tbf}) by assuming $t_B=0$,
and $P=0=Q$, $S= (10^3 + \ell)^{\pm \varepsilon 0.8}$,	
where $+$ is for propagation from the origin
[${  E}'/{  E} =  \varepsilon 0.8 /(10^3 + \ell)$]
and $-$ towards the origin
[${  E}'/{  E} =  - \varepsilon 0.8 /(10^3 + \ell)$].
We consider two models: model 4 with 
$\varepsilon=1$ and model 5 with $\varepsilon=-1$.
As can be seen by using (\ref{rho})--(\ref{Weyl}),
for $r>400$ Mpc these models become almost homogeneous.

\begin{figure*}
\includegraphics[scale=0.8]{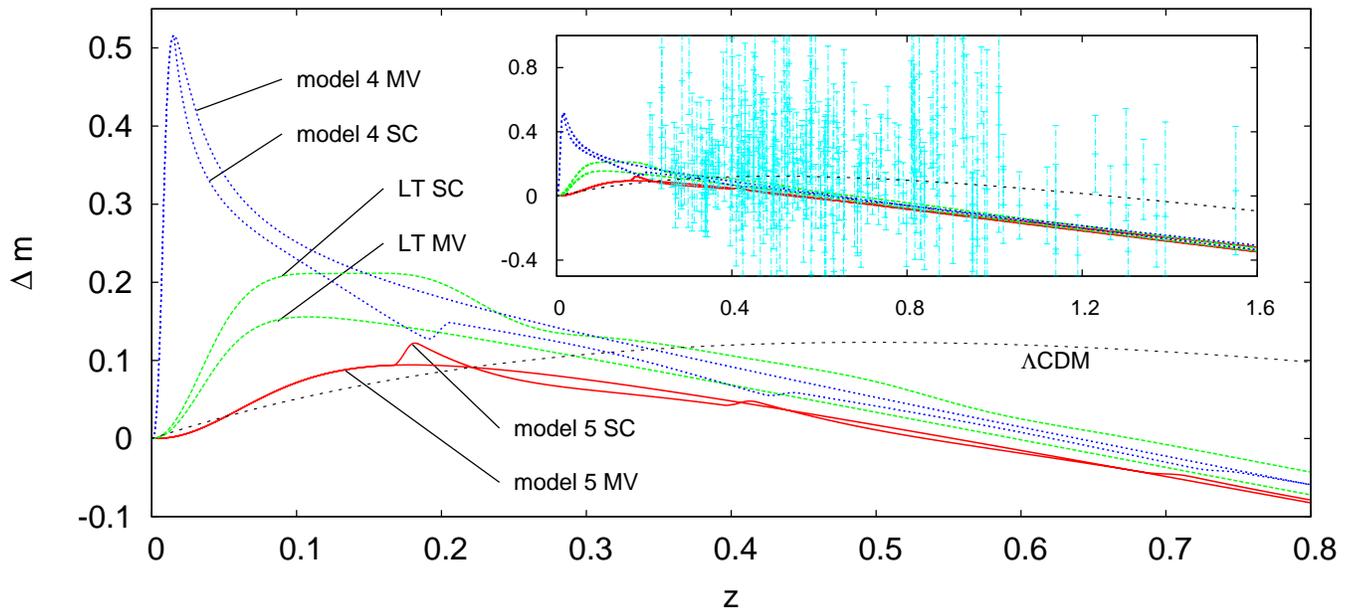}
\caption{The residual Hubble diagram for model 4 (blue, SC -- Swiss Cheese, MV -- minimal void), 
model 5 (red, SC -- Swiss Cheese, MV -- minimal void) and for the corresponding Lema\^itre--Tolman model (i.e. with $M$ and $k$ as in models 4 and 5 but with ${  E}_{,r} = 0$) are also presented (green). 
For clarity, the Union supernova data set \cite{K08} presented in the inset is for $z>0.2$ only.}
\label{fig3}
\end{figure*}

We place the observer at the origin. As in the preceding section we consider two cases: when light starts to propagate from the observer along $E'>0$ (model 4) or along $E'<0$ (model 5).
We join inhomogeneous patches at  $r=400$Mpc. As seen
by construction and from Fig. \ref{fig0} (right panel), this not a perfect matching ---
only the functions $M$ and $k$ are continuous along this boundary but 
not for example the curvature [eqs. (\ref{3dr}) and (\ref{Weyl})].
Thus, in addition to this not perfect matching, we study
the minimal-void scenario,
where there is only one inhomogeneous patch beyond which the model is 
almost homogeneous -- see also  \cite{ABNV09} --- since
there is no matching here the results are exact.
They are presented in Fig. \ref{fig3}.
As in previous sections, the brightness fluctuations at hight
redshifts are very small and thus for $z>0.3$ the differences between
the minimal void scenario (single path) and the Swiss-cheese models
are small. Thus in terms of fitting the supernova data, there
is no big difference whether the minimal void scenario or the Swiss-cheese
model is considered.

Model 4, because of the first peak which omits low-z supernovae,
has very large $\chi^2$, ie. 
$\chi^2=477$ in the Swiss-cheese case and $\chi^2 = 517$ in the minimal void scenario.
The Lema\^itre--Tolman model ($E'=0$) presents a better fit, i.e.
$\chi^2 = 277$ for the Swiss-cheese model and 
$\chi^2 = 279$ in the minimal void scenario.
The best results are obtained in model 5, where
$\chi^2 = 269$ in the Swiss-cheese case and 
$\chi^2 = 269$ in the minimal void scenario.
For the number of degrees of freedom of the sample -- 307 measurements -- this is quite
a good fit, but in comparison with the $\Lambda$CDM model less so.
For this ``concordance'' model we get $\chi^2 = 249$. 
However, if larger structures than $\sim 500$ Mpc are 
considered then one can obtain the same results as in the 
$\Lambda$CDM model (see for example \cite{CBKH09} where
a model with a central Lema\^itre--Tolman overdensity is used to reproduce
$D_L$ with the same form on the observer's past light cone as in the $\Lambda$CDM model;
or \cite{BW09}, the case of the cosmic flow model, where the
density at the current instant is homogeneous and the
best fit model has $\chi^2 = 240$).

\subsection{What if $\Lambda\ne0$?}

So far we considered models with $\Lambda =0$. In this section we would like to show
that if $\Lambda$ is included one can obtain better fits than in the standard case.

\begin{figure*}
\includegraphics[scale=0.8]{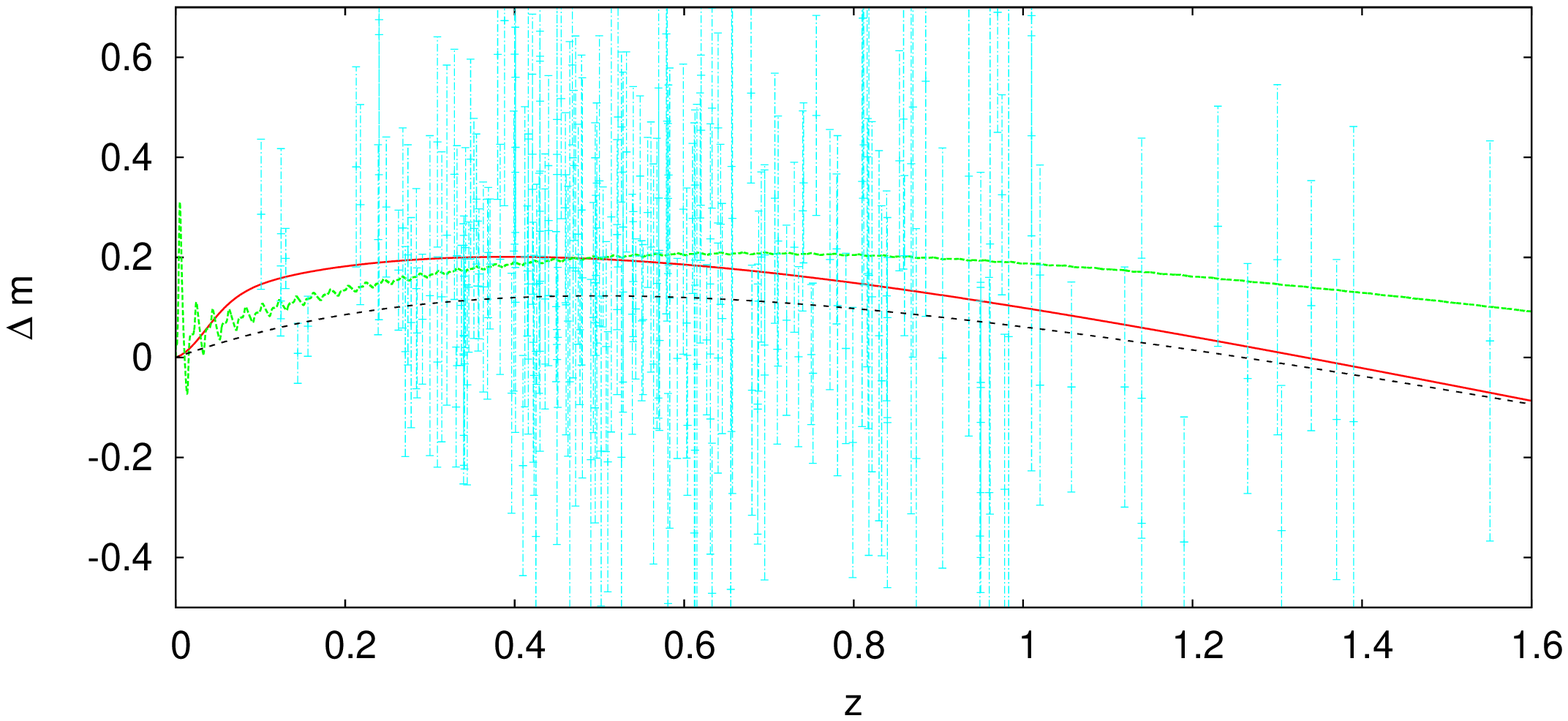}
\caption{The residual Hubble diagram for models 6 (green), 7 (red), and $\Lambda$CDM (dashed line). For clarity the supernova data are presented for $z>0.1$.}
\label{fig4}
\end{figure*}

Let us consider two models: one with small scale density fluctuations: model 6
(which is basically model 1 with $\Omega_m = 0.25$ but in addition we have $\Omega_\Lambda = 0.7$)
and one with large scale density fluctuations: model 7 (in this case a variation of model 5:
but with $\delta_{\rho} = 0.36$, $\Omega_m = 0.25$, $\Omega_\Lambda = 0.6$ and $H_0=72$ km s$^{-1}$ Mpc$^{-1}$).
The residual Hubble diagram is presented in Fig. \ref{fig4}.
For model 6 we have $\chi^2 = 245$ and for model 7, $\chi^2 = 239$ (for the $\Lambda$CDM model $\chi^2 = 249$).
Not only these models have a better $\chi^2$ but also the value of $\Lambda$
changes compared to the standard one. In the case of model 6 the change is at the level of $5\%$,
in the case of model 7 the cosmological constant changes by $15\%$.
This shows that, in the era of precision cosmology, when we want to estimate the
values of the cosmological parameters with high accuracy, we cannot neglect the effect of inhomogeneities.

\section{Conclusions} \label{sec5}

We have used the quasi-spherical axially symmetric subclass of the Szekeres models to construct Swiss-cheese cosmological models which are generalizations of other Swiss-cheeses
where the holes are spherically symmetric and that have already been proposed in the literature. This geometry has already proved to be very useful when studying structure formation as they allow for matter anisotropies \cite{B06,B07}.

However, as regards supernova observations,
we do not get qualitatively different pictures from the ones
obtained with the corresponding Lema\^itre--Tolman Swiss-cheese models.
The results given in Sec. \ref{results} show that,
as in Lema\^itre--Tolman models \cite{BTT07, BTT08, KMM09}, small-scale inhomogeneities ($\sim 50$ Mpc)
do not alter the distance-redshift relation significantly,
and thus cannot fully explain away dark energy.
To reproduce successfully the supernova dimming, we must consider larger structures of order 500 Mpc, as in Lema\^itre--Tolman Swiss-cheeses \cite{ABNV09}.
The presence of such large structures, in the framework of Lema\^itre--Tolman models,
can be ruled out by CMB constraints \cite{WV09}.
However, for Szekeres models, the constraints
might be different since, as shown in \cite{KB09}, the CMB fluctuations are smaller than for the Lema\^itre--Tolman Swiss-cheese models.
 
The fact that in the studied cases we need large patches 
in order to explain supernova dimming without dark energy
might be due to the axial symmetry of the models. Actually, it has been shown, that structure formation is five times faster in some non axially symmetric Szekeres models than in the corresponding Lema\^itre--Tolman models \cite{B06}. Since we know, from \cite{MKMR07}, that the evolution of the inhomogeneities bends more the photon paths compared to the FLRW case than their mere geometry, it should be interesting to investigate if another geometry of the holes might be able to enhance their evolution and therefore to have a stronger effect on the bending of the photon paths and, hence, on the supernova dimming. This will be the subject of future work.


\begin{thebibliography}{10}

\bibitem{celerier00}
M. N. C\'el\'erier, {\it Astron. Astrophys.} {\bf 353}, 63 (2000).

\bibitem{iguchi02}
H. Iguchi, T. Nakamura, and K. Nakao, {\it Prog. Theor. Phys}. {\bf 108}, 809 (2002).

\bibitem{alnes06}
H. Alnes, M. Amarzguioui and \O. Gr\o n, {\it Phys.Rev.} {\bf D73}, 083519 (2006).

\bibitem{chung06}
D. J. H. Chung, and A. E. Romano, {\it Phys Rev}. {\bf D74}, 103507 (2006).

\bibitem{alnes07}
H. Alnes and M. Amarzguioui, {\it Phys. Rev}. {\bf D74} 103520 (2006).

\bibitem{enqvist07} 
K. Enqvist and T. Mattsson, {\it J. Cosmol. Astropart. Phys}. JCAP02(2007)019 (2007).

\bibitem{ABNV09} 
S. Alexander, T. Biswas, A. Notari and D. Vaid, {\it J. Cosmol. Astropart. Phys}. JCAP09(2009)025 (2009).

\bibitem{bolejko08a}
K. Bolejko, {\it PMC Phys.} {\bf A2} 1 (2008); arXiv:astro-ph/0512103

\bibitem{garciabellido08}		
J. Garc\'ia-Bellido and T. Haugb\o lle T, {\it J. Cosmol. Astropart. Phys}. JCAP04(2008)003 (2008).

\bibitem{garciabellido08b}		
J. Garc\'ia-Bellido and T. Haugb\o lle, {\it J. Cosmol. Astropart. Phys}. JCAP09(2008)016 (2008).

\bibitem{zibin08}
J. P. Zibin, A. Moss and D. Scott, {\it Phys.Rev.Lett}. {\bf 101}, 251303 (2008).

\bibitem{yoo08}
C. M. Yoo, T. Kai, and K-i Nakao, {\it Prog. Theor. Phys.} {\bf 120}, 937 (2008).

\bibitem{enqvist08}
K. Enqvist, {\it Gen. Rel. Grav.} {\bf 40}, 451 (2008).

\bibitem{BW09}
K. Bolejko, K. and J. S. B. Wyithe, {\it J. Cosmol. Astropart. Phys} JCAP02(2009)020 (2009).

\bibitem{CBKH09}
M. N. C\'el\'erier, K. Bolejko and A. Krasi\'nski,
{\it Astron. Astrophys.} {\bf 518}, A21 (2010).

\bibitem{tomita00}
K. Tomita, {\it Astrophys. J.} {\bf 529}, 38 (2000).

\bibitem{tomita01a}
K. Tomita, {\it Mon. Not. Roy. Astron. Soc.} {\bf 326}, 287 (2001).

\bibitem{tomita01b}
K. Tomita, {\it Prog. Theor. Phys.} {\bf 106}, 929 (2001).

\bibitem{CS08}
R. R. Caldwell and A. Stebbins, {\it  Phys. Rev. Lett.} {\bf 100} 191302 (2008).

\bibitem{dabrowski98}
M. P. Dabrowski and M. A. Hendry, {\it Astrophys. J}. {\bf 498}, 67 (1998).

\bibitem{godlowski04}
W. God\l{}owski, J. Stelmach and M. Szyd\l{}owski, {\it Class. Quant. Grav}. {\bf 21} 3953 (2004).

\bibitem{stelmach06}
J. Stelmach and I. Jakacka, {\it Class. Quant. Grav.} {\bf 23}, 6621 (2006).

\bibitem{dabrowski07}
M. P. Dabrowski, T. Denkiewicz and M. A. Hendry, {\it  Phys. Rev.} {\bf D75}, 123524 (2007).

\bibitem{MI1} M. Ishak, J. Richardson, D. Garred, D. Whittington, A.
Nwankwo and R. Sussman, {\it Phys. Rev.} {\bf D78}, 123531 (2008).

\bibitem{MI2}
A. Nwankwo, J. Thompson and M. Ishak, to be submitted to Phys. Rev. D. (2010).

\bibitem{Krasinski97}
 A. Krasi\'nski,  {\it Inhomogeneous Cosmological Models},
(Cambridge University Press, Cambridge, 1997).

\bibitem{BTT07}
N. Brouzakis, N. Tetradis and E. Tzavara, {\it J. Cosmol. Astropart. Phys} JCAP02(2007)013 (2007).

\bibitem{MKMR07}
V. Marra, E. W. Kolb, S. Matarrese {\em et al.}, {\it Phys. Rev.} {\bf D76}, 123004 (2007).

\bibitem{BN08}
T. Biswas and A. Notari, {\it J. Cosmol. Astropart. Phys} JCAP06(2008)021 (2008).

\bibitem{BTT08}
N. Brouzakis, N. Tetradis and E. Tzavara, {\it J. Cosmol. Astropart. Phys} JCAP04(2008)008 (2008).

\bibitem{KMM09}
E. W. Kolb, V. Marra and S. Matarrese,
{\it Gen. Relativ. Gravit.} {\bf 42}, 1399 (2010).

\bibitem{CF09}
T. Clifton and P. G. Ferreira, {\it Phys. Rev.} {\bf D80}, 103503 (2009).

\bibitem{B06} 
K. Bolejko, {\it Phys. Rev}. {\bf D73}, 123508 (2006).

\bibitem{B07} 
K. Bolejko, {\it Phys. Rev.} {\bf D75}, 043508 (2007).

\bibitem{S75}
P. Szekeres 1975, {\it Commun. Math. Phys}. {\bf 41}, 55.

\bibitem{H96}
C. Hellaby 1996, {\it Class. Quant. Grav.} {\bf 13}, 2537.

\bibitem{ND07} 
B.C. Nolan and U. Debnath 2007, {\it Phys. Rev.} {\bf D76}, 104046.

\bibitem{BKHC09}
K. Bolejko, A. Krasi\'nski, C. Hellaby and
M.-N. C\'el\'erier, 
\textit{Structures in the Universe by exact methods - formation, evolution,
interactions} (Cambridge University Press, Cambridge, 2009).

\bibitem{KB09}
K. Bolejko, {\it  Gen. Rel. Grav.} {\bf 41}, 1737 (2009).


\bibitem{E71} G.F.R. Ellis 1971, {\it
Proceedings of the International School of Physics `Enrico Fermi', Course 47:
General Relativity and Cosmology}, eds. R. K. Sachs. Academic Press, New
York and London, pp. 104 -- 182; reprinted, with historical comments 2009, in {\it
Gen. Rel. Grav.} {\bf 41}, 581.

\bibitem{S61}
R. K. Sachs 1961, {\it Proc. Roy. Soc. London} {\bf A264}, 309.

\bibitem{SEF92}
P. Schneider, J. Ehlers and E.E. Falco 1992, \textit{Gravitational Lenses}
(Springer-Verlag, New York).

\bibitem{Eth33}
I. M. H. Etherington, {\it Phil. Mag. VII} {\bf 15}, 761 (1933); reprinted, with historical comments  {\it  Gen. Rel. Grav.} {\bf 39}, 1055 (2007).



\bibitem{jon09}
D. H. Jones, M. A. Read, W. Saunders et al., {\it Mon. Not. Roy. Astron. Soc.} {\bf 399}, 683 (2009).

\bibitem{K08}
M. Kowalski et al., {\it Astrophys. J}. {\bf 686}, 749 (2008).


\bibitem{WV09}
W. Valkenburg,  {\it J. Cosmol. Astropart. Phys.} JCAP06(2009)010 (2009).

 \end{thebibliography}
\end{document}